\documentclass{ws-hepph08}

\begin{document}

\newcommand{\ovl}{\overline}
\newcommand{\sT}{{\scriptscriptstyle T}}
\newcommand{\nslash}{\kern 0.2 em n\kern -0.45em /}
\newcommand{\Pslash}{\kern 0.2 em P\kern -0.56em \raisebox{0.3ex}{/}}
\newcommand{\pslash}{\kern 0.2 em p\kern -0.4em /}
\newcommand{\kslash}{\kern 0.2 em k\kern -0.45em /}
\newcommand{\Sslash}{\kern 0.2 em S\kern -0.56em \raisebox{0.3ex}{/}}
\def\adj{{\phantom{h}}}   
\newcommand{\g}{\gamma}
\newcommand{\sig}{\sigma}
\newcommand{\eps}{\epsilon}
\newcommand{\st}{{\scriptscriptstyle T}}
\newcommand{\xbj}{x_{\scriptscriptstyle B}}
\newcommand{\bpt}{\bm p}
\newcommand{\bkt}{\bm k_T}
\newcommand{\bSt}{\bm S}
\newcommand{\ba}{\begin{eqnarray}}
\newcommand{\ea}{\end{eqnarray}}
\newcommand{\beq}{\begin{equation}}
\newcommand{\eeq}{\end{equation}}
\newcommand{\be}{\begin{equation}}
\newcommand{\ee}{\end{equation}}
\newcommand{\bc}{\begin{center}}
\newcommand{\ec}{\end{center}}
\newcommand{\bmi}[1]{\begin{minipage}{#1}}
\newcommand{\emi}{\end{minipage}}
\newcommand{\slsh}[1]{\mbox{$\not\! #1$}}
\newcommand{\ph}{{\rule{0mm}{3mm}}}
\newcommand{\psibar}{\overline{\psi}}
\newcommand{\la}{\langle}
\newcommand{\ra}{\rangle}
\newcommand{\amp}[1]{\la #1 \ra}
\newcommand{\twoamp}[1]{\la \! \la \, #1 \, \ra \! \ra}
\newcommand{\half}{{1\over2}}
\newcommand{\dz}{\int \frac{d^{4}z}{(2\pi)^4}}
\newcommand{\dzp}{\frac{d^{4}z'}{(2\pi)^4}}
\newcommand{\hs}[1]{\hspace{#1}}
\newcommand{\simorder}{\raisebox{-4pt}{$\, \stackrel{\textstyle >}{\sim} \,$}}
\newcommand{\simordertwo}{\raisebox{-3pt}{$\, \stackrel{\textstyle <}{\sim} \,$}}
\newcommand{\Tr}{\text{Tr}}
\newcommand{\open}{{<\kern -0.3 em{\scriptscriptstyle )}}}

\title{\mbox{}\\[-22 mm]
Transversity Asymmetries
\footnote{\uppercase{O}pening talk of the \uppercase{S}econd
\uppercase{I}nternational \uppercase{W}orkshop on
\uppercase{T}ransverse \uppercase{P}olarisation \uppercase{P}henomena in
\uppercase{H}ard \uppercase{P}rocesses (\uppercase{T}ransversity 2008),
\uppercase{F}errara, \uppercase{I}taly,
\uppercase{M}ay 28-31, 2008}
}

\author{Dani\"{e}l Boer}

\address{Department of Physics and Astronomy,\\ 
Vrije Universiteit Amsterdam,\\
De Boelelaan 1081, 1081 HV Amsterdam,\\ 
The Netherlands\\
E-mail: D.Boer@few.vu.nl}

\begin{abstract}
  Ways to access transversity through asymmetry measurements are
  reviewed. The recent first extraction and possible near future
  extractions are discussed.
\end{abstract}


\bodymatter

\section{Transversity from 1978 to 2008}

The year 1978 marks the birth of transversity {\em as a quark 
distribution\/} with the submission  
of the seminal paper \cite{Ralston:1979ys} by Ralston and
Soper on November 14, 1978. 
Transversity was a pre-existing term, but is meant here as
the distribution of transversely polarized quarks inside  
a transversely polarized hadron. 
It depends on the lightcone momentum fraction
$x$ carried by the quark and is often denoted by $h_1(x)$. 
Theoretically it is defined as a hadronic matrix element of a nonlocal
operator:
\beq
\int \frac{d\lambda}{2\pi} e^{i \lambda x} 
{\amp{P,S_T^{}|{\psibar (0) {\cal L}[0, \lambda] 
i \sigma^{i +} \gamma_5 \psi (\lambda n_-)} |P,S_T^{}}} 
= S_T^i \; {h_1(x)},
\eeq
where $P$ and $S_T$ denote the momentum and the transverse spin vector
  of the hadron, $n_-$ is
  a lightlike direction, and
  ${\cal L}$ is a path-ordered exponential that renders the nonlocal
  operator color gauge invariant. 
Transversity is a chiral-odd or helicity-flip quantity, hence, in observables 
it always appears accompanied by another chiral-odd quantity, several
  of which will be discussed below. 
Ralston and Soper considered the double transverse 
spin asymmetry in the Drell-Yan process (reconsidered in detail 
at a later stage in Refs.\
  \cite{Artru:1989zv,Jaffe:1991ra,Cortes:1991ja}), i.e.\ 
the asymmetry in the
azimuthal angular distribution of a produced lepton pair in the collision of
two hadrons, in this case protons, with transverse spins parallel
minus antiparallel: 
\beq
{A_{TT}} = 
\frac{\sigma({p^{\uparrow} \, {p^{\uparrow}}} \to \ell \, \bar \ell \, X) 
\, {\bm{-}} \, \sigma({p^{\uparrow} \, {p^{\downarrow}}} 
\to  \ell \, \bar \ell \, X)}{
\sigma(p^{\uparrow} \, {p^{\uparrow}} \to \ell \, \bar \ell \, X) 
\, {\bm{+}} \, \sigma(p^{\uparrow} \, {p^{\downarrow}} \to  \ell \,
\bar \ell \, X)} \propto \sum_{q} e_q^2\; {h_1^q(x_1) \; 
h{}_1^{\bar q}(x_2)} .
\eeq
However, polarized Drell-Yan is very challenging experimentally, as
witnessed by the fact that even 30 years later it has not yet been
performed. RHIC at BNL is at present the only place that can do double
polarized proton-proton scattering, but $A_{TT}$ is expected to be
small at RHIC. It involves two unrelated transversity functions: the
one for quarks and the one for antiquarks for which likely holds that 
$h_1^{\bar q} \ll h_1^q$. An upper bound on $A_{TT}$ can be obtained
by using Soffer's inequality, 
\beq
{|h_1(x)| \leq \frac{1}{2}
\left[ f_1(x) + g_1(x) \right]}.
\label{Soffer}
\eeq 
The upper bound on {$A_{TT}$} was shown \cite{Martin:1999mg} 
to be {small} at RHIC, of the percent level,
requiring an accuracy that will not be reached soon. 

Already before the advent of RHIC, people started to search for
alternative ways of probing transversity. The first suggestion was
made by Collins \cite{Collins:1992kk}. The idea was to exploit what is
now referred to as the Collins effect, which is parameterized by the
transverse momentum dependent fragmentation function
$H_1^\perp(z,k_T^2)$. It is a spin-orbit coupling effect in the
fragmentation of a transversely polarized quark, resulting in an
asymmetric azimuthal angular distribution of produced hadrons around
the quark polarization direction, a $\sin \phi$ distribution.
$H_1^\perp$ is also a chiral-odd quantity.

Collins pointed out that there would be a $\sin(\phi_h + \phi_S)$ asymmetry 
in semi-inclusive DIS (SIDIS) proportional to $h_1 \otimes
H_1^\perp$. Here $\phi_h$ and $\phi_S$ are the azimuthal angles of 
a final state hadron and the transverse spin of the initial state 
polarized hadron, respectively. The angles are measured w.r.t.\ the
lepton scattering plane, which fixes the polarization state of the
virtual photon such that the helicity flip state of the probed quark 
is selected. 

\begin{figure}
\bc
\psfig{file=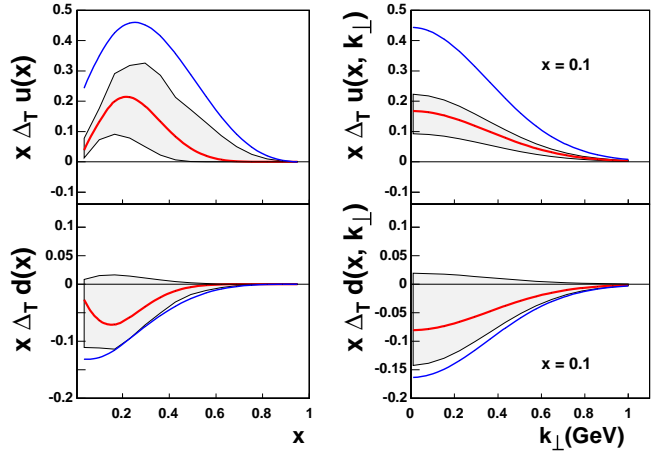,width=2.5in,angle=-90,clip=true}
\caption{First extraction of transversity by Anselmino
{\it et al.}~\cite{Anselmino:2007fs} 
Left: $xh_1(x)$ for $u$ and $d$ quarks, the red curves are the best
fits, the blue curves the Soffer bounds. Right:
$k_T$-dependence using a Gaussian Ansatz.}
\label{firsth1}
\ec
\end{figure}

The HERMES experiment at DESY was the first to
measure a clearly nonzero, percent level, $\sin(\phi_h + \phi_S)$
Collins asymmetry in SIDIS \cite{Airapetian:2004tw,Diefenthaler:2007rj}.
This asymmetry has afterwards also been measured by 
the COMPASS experiment at CERN using a deuteron target 
\cite{Alexakhin:2005iw,Ageev:2006da}
and, as first reported at this Transversity 2008 workshop, 
also on a proton target \cite{Levorato}. These measurements allow for an 
extraction of transversity once the Collins function $H_1^\perp$ is known. 
This exemplifies the crucial role played by electron-positron annihilation
experiments. In Ref.\ \cite{Boer:1997mf} it was pointed out how 
$H_1^\perp$ can be extracted from a $\cos 2 \phi$ asymmetry in 
$e^+ \; e^- \rightarrow \pi^+ \; \pi^- \; X$ (see also the recent, more
extended Ref.\ \cite{Boer:2008fr}).
This turned out to be the method that has actually been employed. 
The measurement of this $\cos 2 \phi$ asymmetry has been performed at KEK using
BELLE data \cite{Seidl:2005,Seidl:2008xc}.
This allowed for the first extraction of transversity by Anselmino
{\it et al.}~\cite{Anselmino:2007fs} in 2007, cf.\ Fig.\ \ref{firsth1}. 

Next we turn to a discussion of the magnitude of the extracted 
transversity functions for $u$ and $d$ quarks.  
Often $h_1$ is compared with its Soffer bound in Eq.\ (\ref{Soffer}) 
or with $g_1$, which is interesting for theoretical reasons, but for 
phenomenology it is more
relevant to compare it to $f_1$, since that is what determines the
magnitude of asymmetries.   
The first extraction, the best fit, indicates that $h_1(x) \approx
f_1(x)/3$, which means that transversity is not particularly small. 
Whether it is of the expected magnitude is a different matter though. 
One way of quantifying this is to compare it to expectations from
lattice QCD and from models for the tensor charge, 
\beq
{\delta q = \int_0^1 dx \left[h_1^q(x)-h_1^{\bar{q}}(x)\right]} ,
\eeq 
which is a {fundamental charge}, like the electric and 
the axial charge. Transversity is the only known way of obtaining the 
tensor charge experimentally.
Using the central fit and assuming antiquark transversity to be small, 
the first extraction yields~\cite{Wakamatsu:2007nc} (at $Q^2 = 2.4$ GeV$^2$)
\[
{\delta u \simeq + 0.39}, \ 
{\delta d \simeq - 0.16}
, \ \text{s.t.}\ {\delta u - \delta d \simeq 0.55}
\]
A lattice determination with two dynamical quark flavors
yields~\cite{Gockeler:2005cj} (at $\mu^2 = 4$ GeV$^2$) 
\[
{\delta u = + 0.857 \pm 0.013}, \ 
{\delta d = - 0.212 \pm 0.005}, \
\text{s.t.} \ {\delta u - \delta d = 1.068 \pm 0.016}
\]
The combination $\delta u - \delta d$ is given, because it has the
advantage of cancelation of 
disconnected contributions which, although expected to be small, are not 
calculated. 

Most models find tensor charges roughly in the range:
\[ {\delta u = + 1.0 \pm 0.2}, \quad 
{\delta d = - 0.2 \pm 0.2}
\]
All of this is consistent with the bounds derived by Soffer 
\cite{Soffer:1994ww}:
\[
|\delta u| \leq 3/2, \quad 
|\delta d| \leq 1/3
\]

The recent
extraction via the Collins effect asymmetries seems to indicate a
$u$-quark tensor charge that is smaller than expected from lattice QCD 
and most models. However, at this workshop we learned that a new fit using
newer and more accurate data yields a larger $\delta u$, which seems 
more in line with expectations \cite{Prokudin}. 

There is nevertheless another issue concerning the magnitude of the extracted
transversity functions. The BELLE and SIDIS data are obtained at different
scales: {$Q^2=110$ GeV$^2$} and {$\amp{Q^2} = 2.4$ GeV$^2$}, respectively. 
The extraction uses two Collins effect asymmetries, which are not like 
ordinary leading twist asymmetries. Both azimuthal asymmetries involve 
transverse momentum dependent functions (TMDs) and beyond tree level this
becomes quite involved. The formalism that deals with TMDs beyond
tree level is that of Collins-Soper (CS) factorization, 
initially considered for 
(almost) back-to-back hadron production in $e^+ e^-$ annihilation 
\cite{Collins:1981uk}{}, and later for SIDIS and Drell-Yan 
\cite{Ji:2004wu,Ji:2004xq}. In principle, 
CS factorization dictates how azimuthal asymmetries depend on $Q^2$, but in
practice this has not been implemented in the $h_1$ extraction
analysis~\cite{Anselmino:2007fs}. Evolution is taken into
account only partially in the following way. 
The Collins function is parameterized in terms 
of the unpolarized fragmentation function $D_1$, 
\beq
H_1^\perp(z,k_T^2) \equiv D_1(z) F(z,k_T^2),
\eeq
and the evolution is taken to be the one of the collinear function
$D_1(z)$. This does not take into
account that beyond tree level also the transverse momentum dependence  
requires modification with changing energy scale. 

\begin{figure}
\bc
\begin{minipage}{4 cm}
\psfig{file=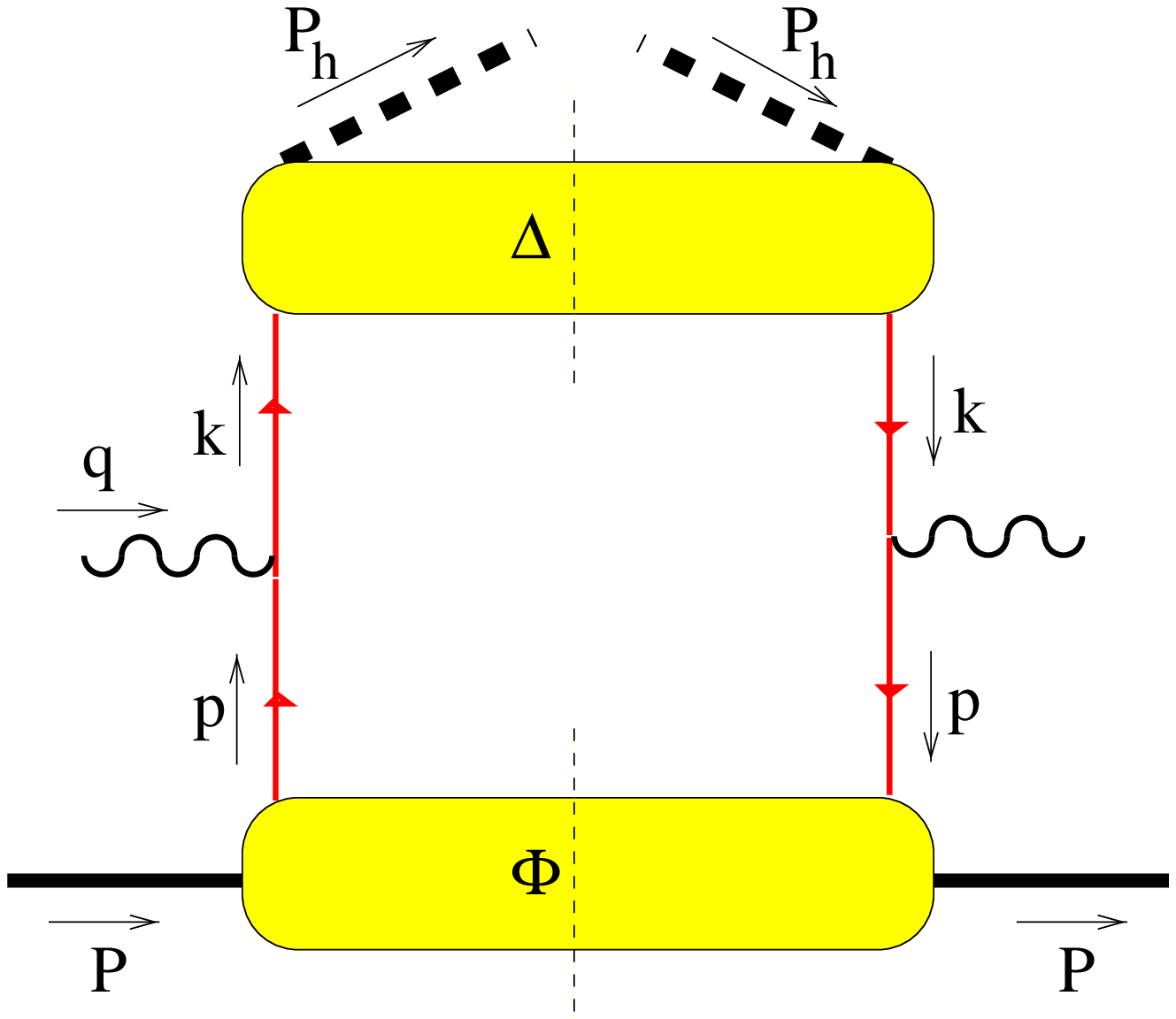,width=1.6in}
\end{minipage}
\hspace{0.5 cm}
\bmi{5 cm}
\psfig{file=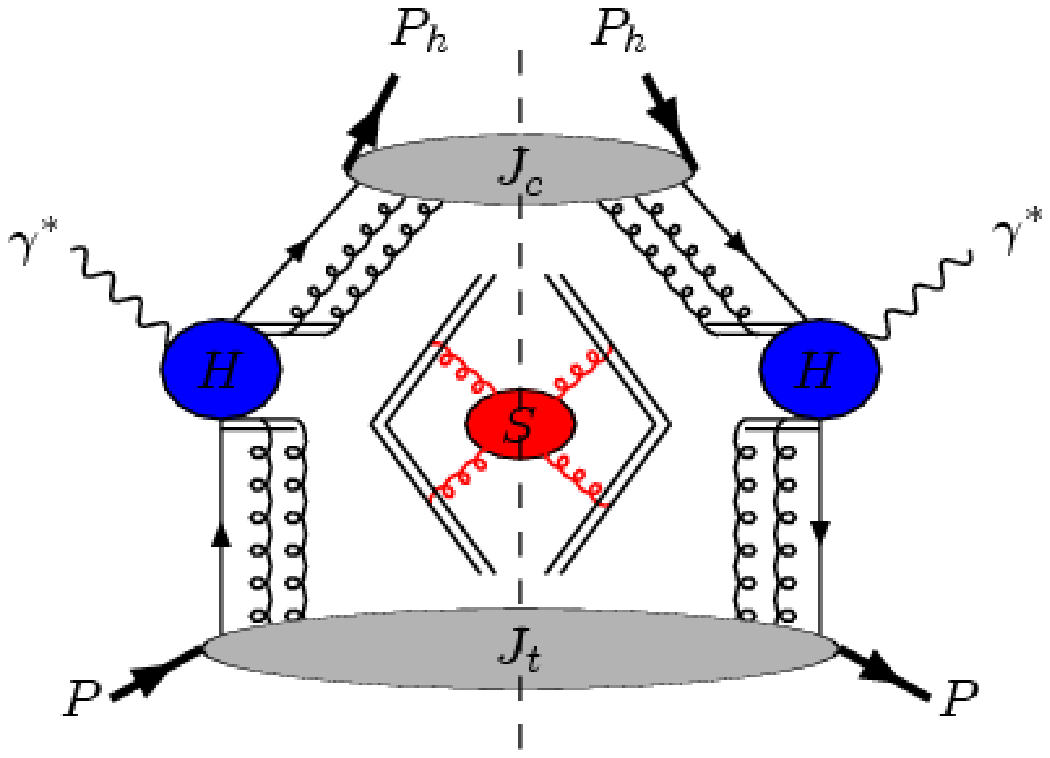,width=2.2in}
\emi
\caption{The left figure shows pictorially the tree level expression ($H=1$ and
$U(l_T^2) \propto \delta(l_T^2)$), the right figure (by F. Yuan) shows the all-order expression (with $S=U$).}
\label{factdiag}
\ec
\end{figure}

Collins effect asymmetries involve {convolution
  integrals}, for example 
the SIDIS asymmetry as a function of the observed
transverse momentum $\bm{q}_T$ (with absolute value $Q_T$), 
\beq
\frac{d\sigma({e \, {p}^\uparrow \to e' \, h \, X})}{
d^{2}\bm{q}_T} \propto\frac{|\bm S_{T}^{}|}{Q_T}\;
{\sin(\phi_h+\phi_{S})}\; 
             {\cal F}\left[\,{\frac{\bm{q}_T\!\cdot \!\bm{k}_T}{M}}\,
                    {h_{1}} {H_1^\perp}\right],
\eeq
involves a convolution that at tree level is of the form:
\ba 
{\cal F}\left[{w}\, {f}\, {D}\right] & \equiv & 
\int d^{2}\bm{p}_T\; d^{2}\bm{k}_T\;
\delta^{(2)} (\bm{p}_T+\bm{q}_T-\bm{k}_T)
\,{w(\bm{p}_T,\bm{q}_T,\bm{k}_T)} \nonumber\\
& & \times 
{f(x,\bm{p}_T^2)} {D(z,z^2\bm{k}_T^2)}. 
\ea
In general however it involves another factor $U$ (called $S$
in Refs.\ \cite{Ji:2004wu,Ji:2004xq}):
\ba 
{\cal F}\left[{w}\, {f}\, {D}\right] & \equiv & 
\int d^{2}\bm{p}_T\; d^{2}\bm{k}_T\; {d^2 \bm{l}_T} \;
\delta^{(2)} (\bm{p}_T+{\bm{l}_T}+\bm{q}_T-\bm{k}_T)
\,{w(\bm{p}_T,\bm{q}_T,\bm{k}_T)} \nonumber \\
& & \times 
{f(x,\bm{p}_T^2)} {D(z,z^2\bm{k}_T^2)} {U({l}_T^{2})}.
\ea
In terms of diagrams the difference is expressed in Fig.\ \ref{factdiag}{}.
As a side remark, we note that it is possible to get rid of the
convolutions by weighted integration over the observed transverse
momentum of the asymmetry. This requires that the asymmetry is
well-described for all values of the transverse momentum, which means
that one has to connect the CS factorization expressions to the
collinear factorization ones that are valid at large transverse momenta. For
the $Q_T$-weighted Collins asymmetry in SIDIS this
works, but for the $Q_T^2$-weighted $\cos 2\phi$ asymmetry in $e^+
e^-$ annihilation a direct extraction of the 
Collins function is not possible in this way 
\cite{Bacchetta:2008xw,Boer:2008fr}.

Beyond tree level the soft factor $U$ dilutes the asymmetry, and
increasingly so as $Q^2$ increases.  Differently stated, for the same
functions $f$ and $D$ and weight $w$, the quantity ${\cal
  F}\left[{w}\, {f}\, {D}\right]$ is smaller beyond tree level. This
effect becomes stronger as $Q$ increases and is referred to as Sudakov
suppression. Conversely, if ${\cal F}\left[{w}\, {f}\, {D}\right]$ is
obtained from experiment and if for instance $f$ is extracted from it
for given $w$ and $D$, then $f$ will be larger when using the
expression beyond tree level.  In Ref.\ \cite{Boer01} this was studied
numerically and a rule of thumb for the $Q^2$ dependence of azimuthal
asymmetries was put forward: asymmetries involving one $k_T$-odd TMD,
such as the Collins effect asymmetry in SIDIS, approximately fall off
as $1/\sqrt{Q}$; asymmetries involving two $k_T$-odd functions, such
as the Collins effect asymmetry in $e^+ e^-$ annihilation,
approximately fall off as $1/Q$.  This behavior was obtained in the
investigated range of $Q=10-100$ GeV and is to a very large extent
independent of model assumptions, even though the magnitude of the
asymmetries does depend heavily on them.

This Sudakov suppression implies that tree level extractions of the Collins
function from the $\cos 2 \phi$ asymmetry at BELLE, leads to an 
underestimation of $H_1^\perp$ (since beyond tree level it will be larger). 
Hence, using that underestimated function to extract
transversity from SIDIS data at a lower $Q^2$ (less Sudakov
suppression), leads to an overestimation of $h_1$. 
Based on the results of Refs.~\cite{Boer01,Boer:2008fr}{}, 
this overestimation may be as large as a factor of 2, 
although there are many uncertainties in this 
estimate and it does not take into account that in Ref.\
\cite{Anselmino:2007fs} some $Q^2$ dependence of $H_1^\perp$ is included 
through the scale dependence of $D_1$, as explained above. 

To get clarity about the magnitude and about the reliability of the
Collins effect extraction method, of course the best would be to do
another independent measurement of transversity. Ideally one wants
this to be a non-TMD, self-sufficient transversity measurement. These
are the cleanest transversity asymmetries that consist of a single
observable that only involves collinear distributions and do not
require experimental input from other experiments done at different
scales and/or using different processes.

Before addressing this topic in detail, it may be worth recalling
that the scale
dependence of $h_1(x)$ itself is quite well-known, i.e.\ to
next-to-leading order   
\cite{Kumano:1997qp,Vogelsang:1997ak,Hayashigaki:1997dn}. The evolution
of $h_1(x,Q^2)$ is very different from that of $g_1(x,Q^2)$, 
in part, because there is no gluon transversity distribution.
$h_1$ grows with increasing $Q^2$ towards smaller $x$, to eventually become
proportional to $\delta(x)$, but with a proportionality constant that
decreases to zero as $Q^2 \to \infty$, hence $h_1(x,Q^2) \to 0$.
Therefore, also the tensor charge decreases with $Q^2$, but it
should be emphasized that it is only very mildly energy scale
dependent. At the Planck scale the
tensor charge is still only reduced by a factor 2 w.r.t.\ $Q^2 =1$ GeV
under next-to-leading order (NLO) evolution.

\section{Transversity asymmetries}

There is an obvious classification of transversity asymmetries into
double and single transverse spin asymmetries, but from a theoretical
point of view there is a more important distinction based on whether
TMDs are involved or not. Cases where collinear factorization can be
applied are much safer than cases for which CS factorization is
expected to apply. The latter usually require some as yet unknown
nonperturbative input and one has to resort to model assumptions, for
instance about the transverse momentum shape of the TMDs, as was done
for the first transversity extraction in Ref.\ 
\cite{Anselmino:2007fs}{}.  For the analysis it also matters whether one
has to combine information from several observables, either obtained
under the same experimental conditions or different ones. Below
these aspects will be discussed for the explicit routes to transversity.

\subsection{Double transverse spin asymmetries}

Almost no experiment aiming to extract $h_1$ will be self-sufficient. 
Most cleanly this requires experiments probing a single ``{$(h_1)^2$}'' 
observable. There are only two such processes: 
\begin{itemize} 
\item $\bar{p}^\uparrow \, p^\uparrow \to \ell \, \bar{\ell} \, X$
 
\item ${p}^\uparrow \, p^\uparrow \to \text{high-}p_T \ \text{jet} \,+\, X $ 
\end{itemize}
Both processes were discussed by Artru and Mekhfi \cite{Artru:1989zv}.
But the first process was only recently considered in detail, 
because of plans to use the future FAIR facility at GSI for its
measurement. The
second process was extensively discussed by Jaffe and Saito
\cite{Jaffe:1996ik}{}, who concluded
that it is likely too challenging to be done at RHIC, because it leads to a 
permille level asymmetry (a result confirmed by Vogelsang
\cite{Vogelsang2000}).  

A somewhat less clean observable is
${\bar{p}^\uparrow \, p^\uparrow \to \pi \, X} $, which is $\propto 
{(h_1)^2} D_1$, considered by Mukherjee, Stratmann and Vogelsang 
\cite{Mukherjee:2005rw}. 
Also ${\bar{p}^\uparrow \, p}$ or 
${\bar{p} \, p^\uparrow}$ Drell-Yan experiments are
self-sufficient, but these involve TMDs and will be discussed in the
next subsection on TMD single spin asymmetries. 

First we look at double transverse spin asymmetries in 
$\bar{p}^\uparrow \, p^\uparrow $ collisions, in particular in 
Drell-Yan. It is ideally suited for $h_1$ extraction,
because $h_1^{\bar q/\bar{p}} = h_1^{q/p}$, leading to:
\beq
{A_{TT}} = 
\frac{\sigma({\bar{p}^{\uparrow} \, {p^{\uparrow}}} \to \ell \,
\bar \ell \, X) \, {\bm{-}} \, \sigma({\bar{p}^{\uparrow} \,
  {p^{\downarrow}}} \to  \ell \, \bar \ell \, X)}{
\sigma(\bar{p}^{\uparrow} \, {p^{\uparrow}} \to \ell \, \bar \ell \, X) 
\, {\bm{+}} \, \sigma(\bar{p}^{\uparrow} \, {p^{\downarrow}} \to
\ell \, \bar \ell \,
X)} \propto \sum_{q} e_q^2\; {h_1^q(x_1) \; 
h{}_1^{q}(x_2)} 
\label{ATTGSI}
\eeq
As said, this can perhaps be done at GSI-FAIR. 
Some of the considered options \cite{Rathmann:2004hr,Nekipelov}
are a collider mode at $\sqrt{s} = 14.5$ GeV
(the currently preferred asymmetric collider option  
of 15 GeV antiprotons on 3.5 GeV protons) and a fixed target mode at 
$\sqrt{s} = 6.7$ GeV (usually quoted as $s=45$ GeV$^2$). 

A chiral quark soliton model calculation~\cite{Efremov:2004qs} of
$h_1$ indicates that large asymmetries of 40-50\% can be expected in
the fixed target mode at $s = 45$ GeV$^2$.  The asymmetry grows with
increasing $Q^2$, but is generally smaller for the higher-$\sqrt{s}$
collider mode. Upper bounds on the asymmetry of approximately 17\% at
$Q=2$ GeV to 38\% at $Q=12$ GeV for the collider mode of $\sqrt{s} =
14.5$ GeV have been obtained by Shimizu {\it et
  al.}~\cite{Shimizu:2005fp}. The first extraction of $h_1$ indicates
that transversity is not much smaller than its upper bound, so
asymmetries of order 10\% at GSI kinematics should be expected. A
Monte Carlo study regarding the feasibility of measuring $A_{TT}$ at
GSI-FAIR is promising \cite{Bianconi:2004wu,Bianconi:2005bd}. But in
the end the success of double polarized Drell-Yan at GSI-FAIR depends
predominantly on whether significant polarization of the antiproton 
beam can be achieved.
   
\mbox{From} the study by Shimizu {\it et al.}~\cite{Shimizu:2005fp} of
the upper bound on $A_{TT}$ it has also become clear that perturbative
corrections hardly affect the asymmetry. The transition from leading
to next-to-leading order pQCD is small and also resummation of large
logs hardly has an effect. A similar robustness can be observed
for a closely related asymmetry investigated in
Ref.~\cite{Kawamura:2008pv}{}, the
unintegrated asymmetry $A_{TT}(Q_T)$, which depends on the transverse
momentum $Q_T$ of the lepton pair w.r.t.\ the beam axis. Although
resummation is essential for this observable, which is described
within the CSS formalism~\cite{Collins:1984kg} that derives from the
CS formalism discussed before, resummation beyond the
leading-logarithmic approximation (LL) has little effect on the
asymmetry. As explained in Ref.\ \cite{Kawamura:2008pv} this is
particular to $\bar{p} \, p$ scattering in the valence region.

In Ref.\ \cite{Kawamura:2008pv} the upper bound of the 
asymmetry $A_{TT}(Q_T)$ for GSI kinematics was shown to be of 
similar magnitude as the integrated asymmetry $A_{TT}$ 
(which is obtained from $A_{TT}(Q_T)$ by integrating 
its numerator and denominator separately). Remarkably, $A_{TT}(Q_T)$ is very 
flat as a function of $Q_T$ and remains flat under $Q^2$ evolution. 

The asymmetry $A_{TT}(Q_T)$ for $p\, p$
scattering~\cite{Kawamura:2007ze},{} which is
considerably smaller for RHIC ($\sqrt{s}=200$ GeV) than J-PARC
($\sqrt{s} = 10$ GeV) kinematics, shows a very different behavior compared to
$\bar{p} \, p$ scattering for potential GSI
kinematics. The asymmetry is flat at LL
level, but not at next-to-leading log. Resummation beyond LL clearly
matters in $p \, p$ collisions.

In conclusion, the double transverse spin asymmetries in 
$\bar{p}^\uparrow \, p^\uparrow$ Drell-Yan 
offer clean, direct and unique probes of transversity and in the 
valence region they are very robust under perturbative corrections. 

As mentioned, one could also consider $A_{TT}^{\pi^0}$ in 
$\bar{p}^\uparrow \, p^\uparrow \to \pi^0 \, X$, which is a 
slightly less clean observable as it requires input on the pion fragmentation
function, which however is quite well-known. Upper bounds for
assumed beam polarizations of 30\% for $\bar p$ and 50\% for $p$
yield asymmetries of a few percent \cite{Mukherjee:2005rw}.
The difference between LO and NLO is a bit larger in this case. 

\subsection{TMD single spin asymmetries}

What if one only has one polarized beam? This question is relevant for
GSI if the antiproton beam cannot be polarized significantly.
For one polarized beam there is a self-sufficient measurement of
transversity which involves TMDs, namely the single spin asymmetry in
{$\bar{p}^\uparrow \, p \to \ell \, \bar{\ell} \, X$} or {$\bar{p} \,
  p^\uparrow \to \ell \, \bar{\ell} \, X$}. Both options are equally
useful, there is no difference theoretically.

\begin{figure}
\bc
\psfig{file=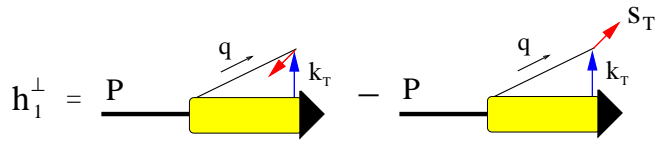,width=3.5in}
\caption{Nonzero $h_1^\perp$ means that the transverse polarization $S_T$
  of quarks (with momentum $q \approx x P+k_T$) inside an unpolarized
  hadron (with momentum $P$) is nonzero. It is a $k_T$-odd and
  chiral-odd TMD.}
\label{h1perp}
\ec
\end{figure}

In the case of one {transversely polarized hadron beam}, there is a 
possible {spin angle $\phi_S$ dependence} of the differential cross section:
\[
\frac{d\sigma}{d\Omega\; d\phi_{S}} \propto 
1+ \lambda \cos^2\theta 
+ \sin^2 \theta \left[ \frac{{\nu}}{2} \; \cos 2\phi - {\rho} \; 
|\bm S_{T}^{}|\;
{\sin(\phi+\phi_{S})} \right] + \ldots 
\]
In a measurement of $\rho$ (from $p^\uparrow - p^\downarrow$) 
also $\nu$ can be
extracted from the same data (from $p^\uparrow + p^\downarrow$), i.e.\ 
under exactly the same experimental conditions. 
This is in contrast to the previously discussed Collins effect
asymmetries. At tree level one has 
\ba
{\nu} \propto {h_1^\perp \, h_1^\perp} & \quad & \text{analogue of
  $\cos 2 \phi$ asymmetry in $e^+e^-$} \nonumber \\
{\rho} \propto {h_1 \, h_1^\perp} & \quad & \text{analogue of
  Collins asymmetry in SIDIS} \nonumber
\ea
These two expressions involve the TMD distribution function~\cite{Boer98} 
$h_1^\perp$, which in some respects is
very similar to the Collins effect fragmentation function, but can be quite
different from it. It is depicted in Fig.\ \ref{h1perp}.

\begin{figure}
\bc
\psfig{file=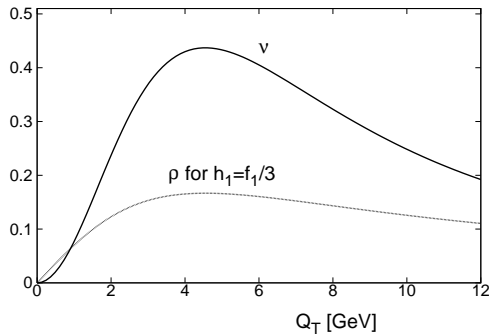,width=2.5in}
\caption{Analyzing power $\nu$ of $\cos 2 \phi$ asymmetry as fitted to NA10
  data using a model Ansatz~\cite{Boer99,Boer99b} for $h_1^\perp$ and 
  the resulting prediction of the single spin asymmetry $\rho$
  using Eq.\ (\ref{rhonu}) for the case $h_1=f_1/3$.}
\label{nurho}
\ec
\end{figure}

The asymmetry {$\nu$} has been measured in  
{$\pi^- \, N \to \mu^+ \mu^- \, X $} by the NA10
Collaboration~\cite{NA10a,NA10b} at CERN 
and the E615 Collaboration~\cite{Conway} 
at Fermilab, roughly 20 years ago. 
The data show an {anomalously large asymmetry}, which differs 
much from the perturbative QCD ${\cal O}(\alpha_s)$ Lam-Tung relation $\nu =
(1-\lambda)/2$ and the ${\cal O}(\alpha_s^2)$ corrections to it 
\cite{BNM93,Mirkes}.
Nonzero $h_1^\perp$ offers an explanation for this discrepancy \cite{Boer99}.
Assuming $u$-quark dominance, Gaussian $k_T^{}$-dependence for 
$h_1^\perp$ and $x$-dependence $\propto f_1(x)$, 
$\rho$ can be related to $\nu$: \cite{Boer99b}
\beq
{\rho} = \frac{1}{2} 
\sqrt{\frac{{\nu} }{{\nu_{\text{max}}}}}\,
{\frac{h_1^u}{f_1^u}}
\label{rhonu}
\eeq
The result is displayed in Fig.\ \ref{nurho}.

The asymmetry $\nu$ for $p\, p$ (e.g.\ at RHIC, where also $\rho$ can be
measured) is expected to be smaller than for $\pi\, p$, due to
absence of valence antiquarks. Preliminary $p\, p$ data from Fermilab were 
shown at this workshop \cite{Peng} and confirm this
expectation. Earlier $p\, d$ data \cite{Zhu:2006gx} also show a small 
asymmetry, probably for the same reason. 

The asymmetry $\nu$ for $\bar{p}\, p$ on the other hand is expected to
be very similar to $\pi\, p$, due to the presence of valence
antiquarks. Therefore, unpolarized $\bar{p} \, p$ Drell-Yan at GSI-FAIR will
likely show a large anomalous $\cos 2\phi$ asymmetry, providing
crucial information about its origin. As explained above, the
measurement of $\nu$ and $\rho$ at GSI-FAIR with only one polarized
beam (either $\bar{p}^\uparrow$ or ${p}^\uparrow$) offers a probe of
transversity.  In this case predominantly $h_1^{\perp u/p}$ and
$h_1^{u/p}$ are accessed, due to the charge-squared factor in Eq.\ 
(\ref{ATTGSI}).

The COMPASS experiment plans to do $\pi^{\pm} \, p^\uparrow$ Drell-Yan
\cite{Colantoni}{}, which although not self-sufficient would provide
valuable information on the flavor dependence of $h_1$ and
$h_1^\perp$.  Especially $\pi^+ p^\uparrow$ is of interest, as there
is no data available on it yet and it provides information on the
$d$-quark ratio $h_1^{\perp d/p}/h_1^{d/p}$, without suppression by a
charge-squared factor. The ratio $\nu/\rho$ for $\pi^{\pm} \,
p^\uparrow$ Drell-Yan in valence approximation namely provides the ratios
$h_1^{\perp u/p}/h_1^{u/p}$ and $h_1^{\perp d/p}/h_1^{d/p}$ for
$\pi^-$ and $\pi^+$, respectively.  Using the input on $h_1^{\perp}$
from for example unpolarized $p \, \bar{p}$ Drell-Yan 
(either from the Tevatron or from GSI-FAIR) would allow for an 
extraction of $h_1$ from $\pi^\pm p^\uparrow$ Drell-Yan at COMPASS. 

The function $h_1^\perp$ may be extracted from $\nu$ at the Tevatron,
but the high $\sqrt{s}$ leads to high $Q^2$ on average. This can
result in considerable Sudakov suppression, which would be
disadvantageous but interesting to verify. One may also probe
$h_1^\perp$ via a $\cos 2 \phi$ asymmetry in photon-jet production $p
\; \bar{p} \to \gamma \; \text{jet}\; X$ at the
Tevatron,~\cite{Boer:2007nd}
\beq
\frac{d\sigma^{h_1 \; h_2\to \gamma \; {\rm jet} \;X}}
{d\eta_\gamma\, d\eta_j \,d^2 \boldsymbol K_{\gamma\perp}\, d^2 \boldsymbol q_{\perp} }   
\propto \left(1 + {\nu_{\text{{\scriptsize DY}}}} \, {R}\, {\cos 2 (\phi_\perp
-\phi_\gamma)}   
 \right)\,
\eeq
where $\phi_\perp$ is the angle of the transverse momentum
$\boldsymbol q_{\perp}$ of the
photon-jet system and $\phi_\gamma$ is the angle of the transverse
momentum $\boldsymbol K_{\gamma\perp}$ of the photon. The analyzing power consists of a proportionality
factor $R$ times $\nu_{\text{{\scriptsize DY}}}$, the $\cos 2\phi$
asymmetry of Drell-Yan probed at the scale 
$|\boldsymbol{K}_{\gamma \perp}|$ which in general is different from
$Q$, which might make a difference from the perspective of Sudakov 
suppression. 
The proportionality factor $R$ is only a function of $f_1$. 
For typical Tevatron kinematics in the central region, recently
investigated for the angular integrated case by the D\O\ Collaboration
\cite{Abazov:2008er}{}, ${\nu_{\text{{\scriptsize DY}}}} \, {R}$ 
was estimated \cite{Boer:2007nd} to be $\sim 5-15\%$. That could be
large enough to allow transversity related TMD studies at the Tevatron too. 

Another ``helper'' process is the ${\cos 2\phi}$ asymmetry $\nu$ in
unpolarized SIDIS $e\, p \to e' \, \pi \, X$, which would be
proportional to $h_1^\perp H_1^\perp$. Given the Collins function it
could in principle be used to extract $h_1^\perp$ too. The asymmetry
in SIDIS turns out to be of quite different size compared to
Drell-Yan. It has been investigated using model calculations in e.g.\ 
Refs.\ \cite{Gamberg:2007wm,Barone:2005kt,Barone:2008tn}{}. The
asymmetries as a function of observed transverse momentum of the pion
are typically on the percent level and are very similar in size for
HERMES kinematics and JLab kinematics (the 12 GeV upgrade).
Interestingly, the $\pi^-$ asymmetries are positive and according to
Ref.\ \cite{Gamberg:2007wm} roughly four times as large as the $\pi^+$
asymmetries which is of opposite sign.  This factor of four is not
related to the charge-squared factor ratio of $u$ and $d$ quarks.

The available data on the ${\cos 2\phi}$ asymmetry in 
unpolarized SIDIS are from EMC and COMPASS; the latter were
  presented at this workshop for the first time \cite{Kafer} 
(soon also HERMES data should become available). The data show that 
$\nu_{\text{{\scriptsize SIDIS}}} \ll \nu_{\text{{\scriptsize DY}}}$. 
The SIDIS data are obtained for not too large values of $Q^2$, where
also higher twist contributions, such as the Cahn effect, can be
relevant. The recent model calculation of Ref.\ \cite{Barone:2008tn}
for instance shows this very clearly. This limits the usefulness of
this observable for transversity related investigations. Nevertheless, it
is interesting to study the importance of higher twist effects at
HERMES and COMPASS energies through this observable. 

It should be added that there is also high $Q^2$ data on the
unpolarized azimuthal asymmetries in SIDIS from ZEUS ($\amp{Q^2}=750$
GeV$^2$). Within the
sizeable errors the ZEUS data are consistent with pQCD 
expectations, but they have
been presented with a lower cut-off on the transverse momentum of the 
final state hadron, cutting out contributions of interest
here. Apart from that, high $Q^2$ is not favorable to probe the $h_1^\perp
H_1^\perp$ contribution due to the Sudakov suppression discussed earlier. 

\subsection{Non-TMD single spin asymmetries}

If one only has one transversely polarized proton beam, then there are
two further possibilities to probe $h_1$ which do not involve
TMDs, i.e.\ to use:
\begin{itemize}
\item transverse $\Lambda$ polarization
\item two hadron systems within a jet 
\end{itemize}
Both options are not self-sufficient, at least
not in a straightforward way; they involve unknown fragmentation
functions, which most cleanly can be obtained from $e^+ e^-$ data. 
The big advantage is though that collinear factorization applies, 
therefore, one only deals with non-TMD functions. 

Transverse $\Lambda$ polarization enters with 
the transversity fragmentation function {$H_1(z)$}. It is still
unknown, but can be measured in $e^+ \; e^- \to 
\Lambda^\uparrow \; \overline{\Lambda}{}^\uparrow X$: $\propto$
{$(H_1)^2$} \cite{Contogouris:1995xc}. 
Subsequently, ${h_1}$ can be accessed via the 
{spin transfer asymmetry} {$D_{NN} \propto h_1 H_1$} in either
$e \; p^\uparrow \to e'\, \Lambda^\uparrow \; X$ or 
$p \; p^\uparrow \to \Lambda^\uparrow \; X$. The latter 
has been measured by the E704 
Collaboration~\cite{Bravar:1997fb}{}, yielding a $D_{NN}$ of order 
20-30\% at a transverse momentum $p_T$ of the $\Lambda$ of around 1
GeV/$c$ ($\sqrt{s} \approx 20$ GeV). However, because of the low
$p_T$, 
this result can probably not be used to
  extract $h_1$ in a trustworthy manner. This should be different at RHIC. 
Upper bounds for $D_{NN}$ calculated \cite{deFlorian:1998am} for
RHIC at $\sqrt{s}=500$ GeV show promisingly large asymmetries 
at much larger $p_T$.

The other option is to use the Interference Fragmentation Function 
$H_1^\open$, or more generally, chiral-odd two-hadron fragmentation functions. 
Consider for definiteness the final state 
{$\vert (\pi^+ \, \pi^-) X \rangle $}, i.e.\ a $\pi^+ \pi^-$
pair inside a jet. The corresponding fragmentation correlation 
function $\Delta(z)$ of this final state can be parameterized as 
\beq
\Delta(z) \propto \left[{D_1} \mbox{$\not\!\! P\,$}
+ i {H_1^{\open}}\,
\frac{{\mbox{$\not\!\! {R}$}_{{T}}}{\mbox{$\not\!\! {P}\,$}}}{2 
M_\pi} \right] ,
\eeq
where the two-hadron fragmentation functions $D_1$ and $H_1^\open$ 
depend on the sum $z$ of the
momentum fractions $z^\pm$ of the $\pi^\pm$ and on
the invariant mass of the two-pion system (not necessarily in a
factorized way as assumed in Ref.\ \cite{Jaffe:1997hf}). 
The momenta appearing are 
${P}= P_{\pi^+} + P_{\pi^-}$ and ${R_T} = (z^+ P_{\pi^-} - z^-
P_{\pi^+})/z$. The $k_T$ of the pion pair w.r.t.\ the fragmentating
  quark is integrated over. See Ref.\ \cite{Bianconi:1999cd} for details.

Nonzero ${H_1^{\open}}$ can arise due to interference between
different partial waves of the $(\pi^+ \, \pi^-)$ system and 
leads to {single spin asymmetries} $ \sin(\phi_{{S_T}}^e +
\phi_{{R_T}}^e)$ in \cite{Ji:1993vw,Collins:1993kq,Jaffe:1997hf}
\ba
{e \, p^\uparrow \rightarrow e' \, (\pi^+ \, \pi^-) \, X} 
& \qquad & \propto h_1 \otimes
{H_1^{\open}} \nonumber \\[1 mm]
{p\, p^\uparrow \to (\pi^+ \, \pi^-) \, X} & \qquad & \propto f_1 
\otimes h_1 \otimes {H_1^{\open}} \nonumber 
\ea
HERMES SIDIS data~\cite{Airapetian:2008sk} below and above the $\rho$ mass
show a nonzero single spin asymmetry (with the same sign), 
which is another indication that
transversity is nonzero. From the comparison~\cite{Bacchetta:2006un} 
of the data to various model predictions for HERMES kinematics using 
different $h_1$ functions, we conclude that the two-hadron asymmetry 
data are compatible with $h_1 \approx f_1/3$, albeit with 
considerable room for other values too. COMPASS data could narrow this
range down.   

As said, both options discussed here have
the advantage that one is dealing
with collinear factorization. This means no Sudakov suppression and no process
dependence. The latter topic will not be addressed here, but is
intimately connected with the gauge invariant definition of TMDs, cf.\
e.g.\ Ref.~\cite{Bomhof:2007xt}{}.
Another advantage is that the evolution equations for $H_1(z)$ and 
$H_1^{\open}(z)$ are known to
next-to-leading order \cite{Stratmann:2001pt,Ceccopieri:2007ip}{}, they 
are in fact the same. Therefore, from a theoretical 
point of view exploiting 
the transversely polarized $\Lambda$ or two-hadron
fragmentation functions currently offers the safest and most
straightforward way to extract transversity.

Like $H_1$, $H_1^{\open}$ can most cleanly be extracted from 
electron-positron annihilation, in this case from 
$e^+ \, e^- \, \to \, (\pi^+ \, \pi^-)_{\text{jet} \, 1} 
\, (\pi^+ \, \pi^-)_{\text{jet} \, 2} \, X$ via a $\cos(
\phi_{{R_{1T}}}^e + \phi_{{R_{2T}}}^e)$ asymmetry \cite{Artru:1995zu}
$\propto (H_1^\open)^{2}$.
Since pions are easier to measure than polarized $\Lambda$'s, this is
probably the easiest route to transversity at this moment.  
BELLE can once again play a
crucial role here (as BABAR could). Its data would allow for a non-TMD 
extraction of transversity
in the not too far future. Therefore, all eyes are on BELLE again in 
this respect.

\section{Routes to transversity}

In the previous section several different routes to transversity were
discussed. They can be classified into four types, which are
summarized in Table \ref{table1} for the 
processes discussed before. On the one hand, there are the
options that use collinear (non-TMD) functions, which are safer from a
theoretical point of view. Some of these options are self-sufficient,
but others require additional input, which most cleanly comes from 
$e^+e^-$ collisions. On the other hand, there are the TMD options,
which are theoretically challenging and it is somewhat ironic to note
that what is theoretically the most challenging option, i.e.\
exploiting the Collins effect, is the one that has been done first
experimentally.     
\begin{table}[h]
\tbl{Summary of routes to transversity}
{\footnotesize
\begin{tabular}{@{}lll@{}}
\hline
{} &{} &{}\\[-1.5ex]
{} & non-TMD & TMD \\[1ex]
\hline
{} &{} &{}\\[-1.5ex]
self-sufficient & {$\bar{p}^\uparrow \, p^\uparrow \to
\ell \, \bar{\ell} \, X$} & {$p \, \bar{p}^\uparrow \to
\ell \, \bar{\ell} \, X$} \\[1ex]
{} & {${p}^\uparrow \, p^\uparrow \to
(\text{high-}p_T \ \text{jet}) \, X $} & {$\bar{p} \,
p^\uparrow \to \ell \, \bar{\ell} \, X$} \\[1ex]
{} &{} &{}\\[-1.5ex]
using external input & {$e \, p^\uparrow \to e'\, \Lambda^\uparrow \, X$} 
& {$e \, p^\uparrow \to e' \, \pi \, X$}\\[1ex]
{} & {$p \, p^\uparrow \to \Lambda^\uparrow \, X$} 
& $\pi \, p^\uparrow \to \ell \, \bar{\ell} \, X$\\[1ex]
{} & {$e \, p^\uparrow \to e' \, (\pi^+ \, \pi^-) \, X$}& \\[1ex] 
{} & {$p\, p^\uparrow \to (\pi^+ \, \pi^-) \, X$} & \\ [1ex]
\hline
\end{tabular}}
\label{table1}
\end{table}

Several remarks have to be added in relation to this table. As pointed
out by Bacchetta and Radici \cite{Bacchetta:2004it} $H_1^\open$ can
also be extracted from $p\, p \to (\pi^+ \, \pi^-) \, (\pi^+ \,
\pi^-)\, X$. Similarly, $H_1$ could be extracted from $p\, p \to
\Lambda^\uparrow \, \bar{\Lambda}^\uparrow \, X$. This makes $p \,
p^\uparrow$ experiments in principle self-sufficient too.  But clearly
this would be less clean and more involved than using $e^+ e^-$
extractions of $H_1^\open$ and $H_1$, due to the appearance of
additional distribution and fragmentation functions, and contributions
from multiple partonic subprocesses. Here, different subprocesses can
enter in numerator and denominator of the asymmetries, because of
which the observables in $p\, p$ are likely to be considerably smaller
than in $e^+ e^-$ annihilation.  Nevertheless, it is important to keep
in mind that one can make the experiments that require a separate
extraction of an unknown fragmentation function, self-sufficient by
considering more complicated $p\, p$ or $e\, p$ processes.  This may
not apply to the Collins function however. It is currently not clear
whether $p\,p \to \pi \, \pi\, X$, where the two pions are in separate
jets, can be used to safely extract the Collins function. Concerns
regarding factorization have been raised in e.g.\ Refs.\ 
\cite{Collins:2007nk,Collins:2007jp}{}.

Note that the process $p\,p^\uparrow \to \pi \, X$ is absent from the
table. This is because the single spin asymmetry $A_N$ is described by
a twist-three expression that consists of several contributions, not
all proportional to transversity. Therefore, it is not clear how to
safely extract it from this observable. Instead, $p\,p^\uparrow \to
\gamma \, \pi \, X$ or $p\,p^\uparrow \to \pi \, \text{jet} \, X$
(cf.\ also Ref.~\cite{Yuan:2008yv,Yuan:2008tv}) could be used,
although also here factorization is yet to be established (which is
the reason for not including them in the table).

Finally, it is worth adding that there is a special role for Drell-Yan
at RHIC: ${p}^\uparrow \, p^\uparrow \to \ell \, \bar{\ell} \, X$. It
offers a clean way to learn about transversity of antiquarks
$h_1^{\bar q}$. Its contribution to the tensor charge is important to
know. It would not be satisfactory to always have to assume that
antiquark transversity is small, without knowing how small. Therefore,
$A_{TT}$ at RHIC is still worth measuring.

\subsection{Transversity GPD}

Now we turn to a completely independent way of accessing transversity
that falls outside the framework of collinear functions and TMDs
discussed thus far. 

There are four chiral-odd Generalized Parton Distributions (GPDs) 
\cite{Diehl:2001pm},
which includes the transversity GPD~\cite{Hoodbhoy:1998vm} $H_T(x,\xi,t)$.
In the forward limit: $H_T(x,0,0) = h_1(x)$, which offers an
alternative way to access transversity and the tensor charge. Of
course, the latter requires not only the extrapolation to the forward
limit, but also integration over all $x$ values. This will be quite
challenging, but it may be worth pursuing this route too because it
can be measured, for instance
at JLab or a future electron-ion collider, 
without the need to polarize the proton. 

Suggestions to probe $H_T$ in {exclusive electroproduction} have been
put forward, for instance, via the production of two vector mesons
in particular polarization states \cite{Ivanov:2002jj,Enberg:2006he}{}, 
$\gamma^* \, p \to \rho_L^0 \, \rho_T^+ \, n$. Very recently it was
suggested \cite{Ahmad:2008hp,Goldstein} 
that transversity could be measured via 
$\gamma^* \, p \to \pi^0 \, p'$. In both cases the idea is that the 
spin states of the photon and the meson(s) enforce
a helicity flip of the quarks inside the proton. In this way there is
no need to polarize the proton. Helicity conservation requires the
helicity flip on the proton side. This is similar to how the axial
charge can be measured in unpolarized elastic $e p$ or $\nu p$
scattering. 

It should be mentioned that in case of single vector meson production, 
e.g.\ $\gamma^* \, p \to \rho_T \, p'$, problems regarding
factorization arise. Unfortunately this process cannot be used to 
extract transversity \cite{Diehl:1998pd,Collins:1999un}. 

Some information on chiral-odd GPDs has already been obtained 
from lattice QCD \cite{Gockeler:2006zu}. Besides yielding 
results for the tensor charge, they also show there to be 
nonzero transverse polarization of quarks inside unpolarized hadrons, 
hinting at nonzero $h_1^\perp$.

\section{Conclusions}

Although transversity is a very difficult quantity to measure, 
several transversity asymmetries have come within reach of present day
and near-future experiments. Thanks to SIDIS data by HERMES and
COMPASS, and $e^+e^-$ annihilation data by BELLE the first extraction of
transversity, exploiting the Collins effect, has been possible. This is
an important step forward. The Collins effect asymmetries involve
${\bm k}_T$-dependent functions, TMDs, and are consequently more
difficult to analyze theoretically. Therefore, an independent, 
preferably non-TMD extraction of transversity is desired. 
For this, ${\bar{p}^\uparrow \, p^\uparrow}$ Drell-Yan would be
the ideal process, but two hadron
fragmentation functions currently offer the most straightforward way.
Many more observables could contribute to our knowledge of
transversity, the tensor charge, and other chiral-odd quantities, such
as $h_1^\perp$. Unpolarized Drell-Yan data and lattice QCD results
strongly suggest that the transverse polarization of quarks inside
unpolarized hadrons, which is encoded by $h_1^\perp$, is 
  nonzero and large. If so, then especially 
${\bar{p}^\uparrow \, p}$ or 
${\bar{p} \, p^\uparrow}$ Drell-Yan offers another promising opportunity to
probe transversity. Perhaps this will be possible at GSI-FAIR. It is a
self-sufficient way of measuring transversity, in the sense that no
information from other experiments or even other processes 
needs to be included in the
analysis. Other, more demanding self-sufficient or nearly
self-sufficient options that do not involve TMDs exist too. 
Amazingly most of these possibilities 
are in principle possible with existing accelerators. Further transversity
measurements are therefore expected in the coming years, contributing
valuably to our understanding of hadron spin. 

\section*{Acknowledgments}

I wish to thank the organizers for their kind invitation to give the 
opening talk at this exciting workshop, where so many new results were
presented. I thank Markus Diehl for a useful 
discussion on vector meson production and Marco Contalbrigo for
some feedback on the text.


\begin{thebibliography}{100}

\bibitem{Ralston:1979ys}
  J.~P.~Ralston, D.~E.~Soper,
  {\em Nucl. Phys. B} {\bf 152}, 109 (1979).

\bibitem{Artru:1989zv}
  X.~Artru, M.~Mekhfi,
  {\em Z. Phys. C} {\bf 45}, 669 (1990).

\bibitem{Jaffe:1991ra}
  R.~L.~Jaffe, X.~Ji,
  {\em Nucl. Phys. B} {\bf 375}, 527 (1992).

\bibitem{Cortes:1991ja}
  J.~L.~Cortes, B.~Pire, J.~P.~Ralston,
  {\em Z. Phys. C} {\bf 55}, 409 (1992).

\bibitem{Martin:1999mg}
  O.~Martin {\it et al.},  
  {\em Phys. Rev. D} {\bf 60}, 117502 (1999).

\bibitem{Collins:1992kk}
  J.~C.~Collins,
  {\em Nucl. Phys. B} {\bf 396}, 161 (1993).

\bibitem{Airapetian:2004tw}
  A.~Airapetian {\it et al.} [HERMES Coll.],
  {\em Phys. Rev. Lett.} {\bf 94}, 012002 (2005).

\bibitem{Diefenthaler:2007rj}
  M.~Diefenthaler  [HERMES Coll.],
  arXiv:0706.2242 [hep-ex].

\bibitem{Alexakhin:2005iw}
  V.~Y.~Alexakhin {\it et al.}  [COMPASS Coll.],
  {\em Phys. Rev. Lett.} {\bf 94}, 202002 (2005).

\bibitem{Ageev:2006da}
  E.~S.~Ageev {\it et al.}  [COMPASS Coll.],
  {\em Nucl. Phys. B} {\bf 765}, 31 (2007).

\bibitem{Levorato}
S. Levorato [COMPASS Coll.], {\em these proceedings.}

\bibitem{Boer:1997mf}
  D.~Boer, R.~Jakob, P.~J.~Mulders,
  {\em Nucl. Phys. B} {\bf 504}, 345 (1997).

\bibitem{Boer:2008fr}
  D.~Boer,
  {\em Nucl. Phys. B} (2008) {\em in press}, arXiv:0804.2408 [hep-ph].

\bibitem{Seidl:2005}
  R.~Seidl {\it et al.}  [BELLE Coll.],
  {\em Phys. Rev. Lett.} {\bf 96}, 232002 (2006).

\bibitem{Seidl:2008xc}
  R.~Seidl {\it et al.}  [BELLE Coll.],
  arXiv:0805.2975 [hep-ex].

\bibitem{Anselmino:2007fs}
  M.~Anselmino {\it et al.}, 
  {\em Phys. Rev. D} {\bf 75}, 054032 (2007).

\bibitem{Wakamatsu:2007nc}
  M.~Wakamatsu,
  {\em Phys. Lett. B} {\bf 653}, 398 (2007).

\bibitem{Gockeler:2005cj}
  M.~G\"ockeler, Ph.~H\"agler {\it et al.}  [QCDSF Coll.],
  {\em Phys. Lett. B} {\bf 627}, 113 (2005).

\bibitem{Soffer:1994ww}
  J.~Soffer,
  {\em Phys. Rev. Lett.} {\bf 74}, 1292 (1995).

\bibitem{Prokudin}
A. Prokudin, {\em these proceedings.}

\bibitem{Collins:1981uk}
  J.~C.~Collins, D.~E.~Soper,
  {\em Nucl. Phys. B} {\bf 193}, 381 (1981).

\bibitem{Ji:2004wu}
  X.~Ji, J.-P.~Ma, F.~Yuan,
  {\em Phys. Rev. D} {\bf 71}, 034005 (2005).

\bibitem{Ji:2004xq}
  X.~Ji, J.-P.~Ma, F.~Yuan,
  {\em Phys. Lett. B} {\bf 597}, 299 (2004).

\bibitem{Bacchetta:2008xw}
  A.~Bacchetta, D.~Boer, M.~Diehl, P.~J.~Mulders,
  {\em JHEP} {\bf 0808}, 023 (2008).

\bibitem{Boer01}
D.~Boer,
{\em Nucl.\ Phys. B} {\bf 603}, 195 (2001).

\bibitem{Kumano:1997qp}
  S.~Kumano, M.~Miyama,
  {\em Phys. Rev. D} {\bf 56}, 2504 (1997).

\bibitem{Vogelsang:1997ak}
  W.~Vogelsang,
  {\em Phys. Rev. D} {\bf 57}, 1886 (1998).

\bibitem{Hayashigaki:1997dn}
  A.~Hayashigaki, Y.~Kanazawa, Y.~Koike,
  {\em Phys. Rev. D} {\bf 56}, 7350 (1997).

\bibitem{Jaffe:1996ik}
  R.~L.~Jaffe, N.~Saito,
  {\em Phys. Lett. B} {\bf 382}, 165 (1996).

\bibitem{Vogelsang2000}
W. Vogelsang, talk presented at the RBRC Workshop on ``Future Transversity
Measurements'' held at BNL, September 18-20, 2000.

\bibitem{Mukherjee:2005rw}
  A.~Mukherjee, M.~Stratmann, W.~Vogelsang,
  {\em Phys. Rev. D} {\bf 72}, 034011 (2005).

\bibitem{Rathmann:2004hr}
  F.~Rathmann, P.~Lenisa,
  arXiv:hep-ex/0412078.

\bibitem{Nekipelov}
M. Nekipelov, {\em these proceedings.}

\bibitem{Efremov:2004qs}
  A.~V.~Efremov, K.~Goeke, P.~Schweitzer,
  {\em Eur. Phys. J. C} {\bf 35}, 207 (2004).

\bibitem{Shimizu:2005fp}
  H.~Shimizu {\it et al.}, 
  {\em Phys. Rev. D} {\bf 71}, 114007 (2005).

\bibitem{Bianconi:2004wu}
  A.~Bianconi, M.~Radici,
  {\em Phys. Rev. D} {\bf 71}, 074014 (2005).

\bibitem{Bianconi:2005bd}
  A.~Bianconi, M.~Radici,
  {\em Phys. Rev. D} {\bf 72}, 074013 (2005).

\bibitem{Kawamura:2008pv}
  H.~Kawamura, J.~Kodaira, K.~Tanaka,
  {\em Phys. Lett. B} {\bf 662}, 139 (2008).

\bibitem{Collins:1984kg}
  J.~C.~Collins, D.~E.~Soper and G.~Sterman,
  {\em Nucl. Phys. B} {\bf 250}, 199 (1985).

\bibitem{Kawamura:2007ze}
  H.~Kawamura, J.~Kodaira, K.~Tanaka,
  {\em Nucl. Phys. B} {\bf 777}, 203 (2007).

\bibitem{Boer98}
D.~Boer, P.~J.~Mulders,
{\em Phys. Rev. D} {\bf 57}, 5780 (1998).

\bibitem{NA10a}
S.~Falciano {\it et al.} [NA10 Coll.], 
{\em Z. Phys. C} {\bf 31}, 513 (1986).

\bibitem{NA10b}
M.~Guanziroli {\it et al.} [NA10 Coll.],
{\em Z. Phys. C} {\bf 37}, 545 (1988).

\bibitem{Conway}
J.~S.~Conway {\it et al.},
{\em Phys. Rev. D} {\bf 39}, 92 (1989).

\bibitem{BNM93}
A.~Brandenburg, O.~Nachtmann, E.~Mirkes,
{\em Z. Phys. C} {\bf 60}, 697 (1993).

\bibitem{Mirkes} 
E.~Mirkes, J.~Ohnemus,
{\em Phys. Rev. D} {\bf 51}, 4891 (1995).

\bibitem{Boer99}
D.~Boer,
{\em Phys. Rev. D} {\bf 60}, 014012 (1999).

\bibitem{Boer99b}
D.~Boer,
{\em Nucl. Phys. B Proc. Suppl.}  {\bf 79}, 638 (1999).

\bibitem{Peng}
J.C. Peng [FNAL-E866/NuSea Coll.], {\em these proceedings.}

\bibitem{Zhu:2006gx}
  L.~Y.~Zhu {\it et al.}  [FNAL-E866/NuSea Coll.],
  {\em Phys. Rev. Lett.} {\bf 99}, 082301 (2007).

\bibitem{Colantoni}
M. Colantoni [COMPASS Coll.], {\em these proceedings.}

\bibitem{Gamberg:2007wm}
  L.~P.~Gamberg, G.~R.~Goldstein, M.~Schlegel,
  {\em Phys. Rev. D} {\bf 77}, 094016 (2008).

\bibitem{Barone:2005kt}
  V.~Barone, Z.~Lu, B.~Q.~Ma,
  {\em Phys. Lett. B} {\bf 632}, 277 (2006).

\bibitem{Barone:2008tn}
  V.~Barone, A.~Prokudin, B.~Q.~Ma,
  arXiv:0804.3024 [hep-ph].

\bibitem{Kafer}
W. Kafer [COMPASS Coll.], {\em these proceedings.}

\bibitem{Boer:2007nd}
  D.~Boer, P.~J.~Mulders, C.~Pisano,
  {\em Phys. Lett. B} {\bf 660}, 360 (2008).

\bibitem{Abazov:2008er}
  V.~M.~Abazov {\it et al.}  [D\O\ Coll.],
  arXiv:0804.1107 [hep-ex].

\bibitem{Contogouris:1995xc}
  A.~P.~Contogouris {\it et al.}, 
  {\em Phys. Lett. B} {\bf 344}, 370 (1995).

\bibitem{Bravar:1997fb}
  A.~Bravar {\it et al.}  [E704 Coll.],
  {\em Phys. Rev. Lett.} {\bf 78}, 4003 (1997).

\bibitem{deFlorian:1998am}
  D.~de Florian {\it et al.}, 
  {\em Phys. Lett. B} {\bf 439}, 176 (1998).

\bibitem{Jaffe:1997hf}
  R.~L.~Jaffe, X.~Jin, J.~Tang,
  {\em Phys. Rev. Lett.} {\bf 80}, 1166 (1998).

\bibitem{Bianconi:1999cd}
  A.~Bianconi, S.~Boffi, R.~Jakob, M.~Radici,
  {\em Phys. Rev. D} {\bf 62}, 034008 (2000).

\bibitem{Ji:1993vw}
  X.~Ji,
  {\em Phys. Rev. D} {\bf 49}, 114 (1994).

\bibitem{Collins:1993kq}
  J.~C.~Collins, S.~F.~Heppelmann, G.~A.~Ladinsky,
  {\em Nucl. Phys. B} {\bf 420}, 565 (1994).

\bibitem{Airapetian:2008sk}
  A.~Airapetian {\it et al.}  [HERMES Coll.],
  {\em JHEP} {\bf 0806}, 017 (2008). 

\bibitem{Bacchetta:2006un}
  A.~Bacchetta, M.~Radici,
  {\em Phys. Rev. D} {\bf 74}, 114007 (2006).

\bibitem{Bomhof:2007xt}
  C.~J.~Bomhof, P.~J.~Mulders,
  {\em Nucl. Phys. B} {\bf 795}, 409 (2008).

\bibitem{Stratmann:2001pt}
  M.~Stratmann, W.~Vogelsang,
  {\em Phys. Rev. D} {\bf 65}, 057502 (2002).

\bibitem{Ceccopieri:2007ip}
  F.~A.~Ceccopieri, M.~Radici, A.~Bacchetta,
  {\em Phys. Lett. B} {\bf 650}, 81 (2007).

\bibitem{Artru:1995zu}
  X.~Artru, J.~C.~Collins,
  {\em Z. Phys. C} {\bf 69}, 277 (1996).

\bibitem{Bacchetta:2004it}
  A.~Bacchetta, M.~Radici,
  {\em Phys. Rev. D} {\bf 70}, 094032 (2004).

\bibitem{Collins:2007nk}
  J.~Collins, J.~W.~Qiu,
  {\em Phys. Rev. D} {\bf 75}, 114014 (2007).

\bibitem{Collins:2007jp}
  J.~Collins,
  arXiv:0708.4410 [hep-ph].

\bibitem{Yuan:2008yv}
  F.~Yuan,
  {\em Phys. Rev. D} {\bf 77}, 074019 (2008).

\bibitem{Yuan:2008tv}
  F.~Yuan,
  {\em Phys. Lett. B} {\bf 666}, 44 (2008).

\bibitem{Diehl:2001pm}
  M.~Diehl,
  {\em Eur. Phys. J. C} {\bf 19}, 485 (2001).

\bibitem{Hoodbhoy:1998vm}
  P.~Hoodbhoy, X.~Ji,
  {\em Phys. Rev. D} {\bf 58}, 054006 (1998).

\bibitem{Ivanov:2002jj}
  D.~Y.~Ivanov {\it et al.}, 
  {\em Phys. Lett. B} {\bf 550}, 65 (2002).

\bibitem{Enberg:2006he}
  R.~Enberg, B.~Pire, L.~Szymanowski,
  {\em Eur. Phys. J. C} {\bf 47}, 87 (2006).

\bibitem{Ahmad:2008hp}
  S.~Ahmad, G.~R.~Goldstein, S.~Liuti,
  arXiv:0805.3568 [hep-ph].

\bibitem{Goldstein}
G.~R.~Goldstein, {\em these proceedings.}

\bibitem{Diehl:1998pd}
  M.~Diehl, T.~Gousset, B.~Pire,
  {\em Phys. Rev. D} {\bf 59}, 034023 (1999).

\bibitem{Collins:1999un}
  J.~C.~Collins, M.~Diehl,
  {\em Phys. Rev. D} {\bf 61}, 114015 (2000).

\bibitem{Gockeler:2006zu}
  M.~G\"ockeler {\it et al.}  [QCDSF Coll.],
  {\em Phys. Rev. Lett.}  {\bf 98}, 222001 (2007).

\end{thebibliography}
\end{document}